\documentclass[aps,reprint,superscriptaddress]{revtex4-1}

\usepackage{graphicx}% Include figure files
\usepackage{dcolumn}% Align table columns on decimal point
\usepackage{bm}% bold math
\usepackage{mathrsfs}
\usepackage[colorlinks]{hyperref}
\usepackage[T1]{fontenc}
\usepackage[utf8]{inputenx}
\usepackage{amsmath}
\usepackage{verbatim}
\usepackage{graphicx}
\usepackage{xcolor}
\usepackage{appendix}

\begin{document}

\title{Precursors to Rare Events in Stochastic Resonance}
%\shorttitle{Title} %Insert here a short version of the title if it exceeds 70 characters

 \author{L. T. Giorgini}
\email{ludovico.giorgini@su.se}
\affiliation{Nordita, Royal Institute of Technology and Stockholm University, Stockholm 106 91, Sweden}

\author{S.H. Lim}
\affiliation{Nordita, Royal Institute of Technology and Stockholm University, Stockholm 106 91, Sweden}

\author{W. Moon}
\affiliation{Department of Mathematics, Stockholm University 106 91 Stockholm, Sweden}
\affiliation{Nordita, Royal Institute of Technology and Stockholm University, Stockholm 106 91, Sweden}

\author{J.S. Wettlaufer}
\affiliation{Yale University, New Haven, Connecticut 06520, USA}
%\affiliation{Mathematical Institute, University of Oxford, Oxford OX2 6GG, UK}
\affiliation{Nordita, Royal Institute of Technology and Stockholm University, Stockholm 106 91, Sweden}

\date{\today}

\begin{abstract}
In stochastic resonance, a periodically forced Brownian particle in a double-well potential jumps between minima at rare increments, the prediction of which poses a major theoretical challenge.  Here, we use a path-integral method to find a precursor to these transitions by determining the most probable (or ``{optimal}'') space-time path of a particle.  We characterize the optimal path using a direct comparison principle between the Langevin and Hamiltonian dynamical descriptions, allowing us to express the jump condition in terms of the accumulation of noise around the stable periodic path.  In consequence, as a system approaches a rare event these fluctuations approach one of the deterministic minimizers, thereby providing a precursor for predicting a stochastic transition. We demonstrate the method numerically, which allows us to determine whether a state is following a stable periodic path or will experience an incipient jump with a high probability. The vast range of systems that exhibit stochastic resonance behavior insures broad relevance of our framework, which allows one to extract precursor fluctuations from data.
\end{abstract}

\maketitle

\section{1. Introduction}
Rare events, which frequently accompany fluctuations or phase transitions, arise in a wide range of natural and social systems, such as infectious disease outbreaks, earthquakes, stock market crashes, and many others \cite[e.g.][]{ghil2011extreme, forgoston2018primer,farazmand2018extreme}. 
Of particular interest are dynamical systems that have bifurcations, at which sudden transitions 
to distinct dynamical regimes occur \cite{scheffer2009early,dematteis2019}. Even before reaching a bifurcation, noise-induced transitions can occur with low probability \cite{hanggi1990reaction}.
In consequence, a system experiences a large-magnitude change
resulting in significant positive or deleterious consequences. 
Hence, it is important to understand the mechanism leading to the occurrence of such events, and to seek precursors to anticipate them \cite{faranda2014}. 

The desire to predict these rare events in advance has fueled studies, to simulate \cite{grafke2019numerical}, classify \cite{ashwin2012tipping}, analyze \cite{romano2018analysis} and predict \cite{ritchie2016early, kuehn2018early} their properties. Although the existence of early-warning signals for rare events  has been suggested, there are few results determining reliable and robust indicators for noise-induced transitions \cite{chen2019noise}. Because most systems are inherently noisy, understanding the role of noise in inducing these transitions is critical for their quantitative prediction well in advance.
Here we describe a theory quantifying the role of noise in rare events, which underlies probabilistic forecast models. 

We study noise-induced transitions using a class of periodically forced low dimensional stochastic dynamical systems 
and identify a novel early-warning indicator for the jumps from one stable state  of the system to another. Periodically forced 
stochastic systems are ubiquitous in nature. For example, periodic forcing and background noise are the main ingredients 
of stochastic resonance \cite{benzi1981mechanism,benzi1982stochastic,herrmann2013stochastic} (see \cite{gammaitoni1998stochastic,wellens2003stochastic} for reviews), wherein the response to a weak signal is magnified by noise induced fluctuations that drive hopping from one stable state to the other in a double-well potential with two minima. Settings of relevance range from the human cardiovascular system \cite{stefanovska1999physics}  to the seasonal variability of the Earth's climate \cite{moon2017unified}. 

This paper is organized as follows. In \S 2, we provide an outline of the mathematical formulation, with the details provided in the Supplementary Material. In \S 3, we discuss the task of finding precursors for the occurrence of a rare event.
We propose a data-driven strategy to study the problem in \S 4.  This strategy constitutes the main contribution of our paper and is presented as a five-step procedure. 
We test this strategy with an example in \S 5 using two different numerical simulations before concluding in \S 6.

\label{sect_outline}
\section{2. Outline of the Mathematical Formulation}

In order to insure our treatment is reasonably self-contained, here we outline the principal waypoints of the path-integral treatment of stochastic processes. For readers not intimate with this approach we have provided details in the Supplementary Material.

The state or position, $x$, of the system is modeled by the following nonautonomous one-dimensional overdamped Langevin equation:
\begin{equation}
\dot{x}(t)=F(x(t),t)+\sqrt{2}\sigma \xi(t),
\label{lang}
\end{equation}
in which 
\begin{equation}
F(x,t)=-U'(x)+A\cos(\omega t),
\end{equation}
where the dot (prime) denotes differentiation with respect to time (position), $U(x)$ is a multi-well potential, $A\cos(\omega t)$ is external periodic forcing and $\xi(t)$ is zero mean Gaussian white noise with correlation function
\begin{equation}
\langle \xi(t)\xi(s)\rangle=\delta(t-s).
\label{noise_corr}
\end{equation}	
We study systems described by Eqs. \eqref{lang}-\eqref{noise_corr} with the (constant) noise intensity $ \sigma \ll A $ by employing a path integral formulation \cite{onsager1953fluctuations}.
The use of this formulation allows us to identify the most probable (optimal) trajectories (also called instantons) among all the possible trajectories that the system state follows to go from a point with the space-time coordinates $(x_i,t_i)$ to another point with the coordinates $(x_f,t_f)$. These optimal paths can be derived by studying large deviations from the unperturbed deterministic dynamics of the system in the weak noise regime (see \cite{freidlin2012random} for the details of sample-path large deviation theory for stochastic differential equations). 

We are interested in the behavior of the system {\it shortly before} its state jumps from one potential well to another. Since $t_f-t_i$ is finite, there exists a finite number of optimal paths, of the order of $(t_f-t_i)/T$, where $T=2\pi/\omega$ is the period of the external periodic forcing.
In fact, it can be shown that these optimal paths, denoted $ x_k(t) $ (with the subscript $k$ denoting a particular path), satisfy the following system of first order differential equations \cite{lehmann2000surmounting}
\begin{align} 
\dot{x}_k(t) &=2p_k(t)+F(x_k(t),t),  \label{x} \\
\dot{p}_k(t) &=-p_k(t)F'(x_k(t),t), 
\label{xp} 
\end{align}
with the boundary conditions 
\begin{equation}
x_k(t_i)=x_i \qquad \text{and} \qquad x_k(t_f)=x_f.
\label{boundary}
\end{equation}
We have introduced the conjugate momenta $ p_k(t) $ relative to the optimal paths $ x_k(t)$. These momenta are defined as $p_k(t) := \frac{1}{2}[\dot{x}_k(t) - F(x_k(t),t)]$ and they measure the deviation from the deterministic unperturbed dynamics. 
Each path $ x_k(t) $ starts at $ t=t_i $ and first follows a stable periodic orbit $ x_s(t) $, defined as the solution of Eq. (\ref{lang}) with $ \sigma=0 $:  
\begin{equation}
\dot{x}_s(t) =F(x_s(t),t) \qquad \text{and} \qquad x_s(t)=x_s(t+T).
\label{no_noise}
\end{equation}
The path begins to deviate from this periodic orbit at a random time $t_0$ and then transitions to a path that closely follows another stable periodic orbit.
This random time $t_0$, which also denotes the time at which the $p_k(t)$ begin to deviate from zero, differs for different realizations of system paths described by Eq. (\ref{lang}).

Formally, the probability distribution that the process $x(t)$ reaches a point $x_f$ at time $t_f$, given 
that it started at a point $x_i$ at time $t_i$ can be written as 
\begin{equation} 
P(x_f,t_f|x_i,t_i)= \sum_{k=1}^{n} P_k(x_f,t_f|x_i,t_i), \label{probb}
\end{equation} 
with $ n= \lfloor{(t_f-t_i)/T}\rfloor $, where $\lfloor \cdot \rfloor$ denotes the floor function.
Each optimal path $x_k$ gives the contribution 
\begin{equation}
P_k(x_f,t_f|x_i,t_i) = \dfrac{1}{\sqrt{4\pi\sigma^2 Q_k(t_f)}}e^{-S[x_k]/\sigma^2}
\label{probbb}
\end{equation}
to the series defining Eq. \eqref{probb}.  Here
\begin{equation}
S[x_k(t)]=\int_{t_i}^{t_f}p_k^2(t) dt
\label{S}
\end{equation}
where the $Q_k$ satisfy the following second order initial value problem \cite{lehmann2000surmounting}:
\begin{equation}
\frac{\ddot{Q}_k(t)}{2}-\partial_t[Q_k(t)F'(x_k(t),t)]+Q_k(t)p_k(t)F''(x_k(t),t)=0,
\label{Q}
\end{equation} 
with
\begin{equation}
Q_k(t_i)=0, \;\;\;\;\; \dot{Q}_k(t_i)=1.
\end{equation}

\section{3. The early-warning indicator}
\label{sect_ewi}

For a long time interval with high probability the system will follow a stable periodic orbit, $ x_s(t) $, around one of the local minima of the potential with fluctuations of order $ \sigma $. 
Rarely, however, the system will jump from one minimum to the other, in which case the most probable path is described by Eqs.  \eqref{x}-\eqref{boundary}, with the jump beginning at time $t_i$ and ending at time $t_f$.  Our principal goal is to obtain quantitative precursors of such rare events and, combined with knowledge of the optimal path, estimate the most probable time of the rare event.

The key observation is as follows.  We compare Eq. (\ref{x}) with the Langevin equation (\ref{lang}) and observe that the optimal condition for the system to jump from one potential well to the other is when the fluctuations around the stable periodic path,
\begin{equation}
\xi(t)= \frac{1}{\sqrt{2} \sigma}[\dot{x}(t)-F(x(t),t)],
\label{noise_acc}
\end{equation} 
accumulate to $\sqrt{2} p(t)/\sigma$, where $p(t)$ is one of the $p_k(t)$'s satisfying Eq. \eqref{xp}. Namely, up to a multiplicative factor of $\sqrt{2}/\sigma$, as the system approaches a rare event, the fluctuations around the stable state approach one of the deterministic minimizers $\sqrt{2} p_k(t)/\sigma$. Therefore, it is crucial to extract such fluctuations from data in order to determine whether a state is simply following a stable periodic path, or begins to follow Eqs. \eqref{x}-\eqref{boundary}, describing the most probable path that can lead the system to jump. 

Clearly, $p(t)$ acts as a forcing for $x(t)$ and hence the former ``anticipates'' the latter.  Thus, although when $p\gg\sigma $ and $|x-x_s|\gg \sigma$ the instanton and its conjugate momentum can be resolved, the former condition is satisfied before the latter condition.  Therefore, $p(t)$ is a better early warning indicator than $x(t)$.  Now, despite the momentum being observable when $p\gg\sigma$, the influence of the oscillatory forcing term with amplitude $A$ in Eq. \eqref{x} is to delay the effect of $p(t)$ on $x(t)$
until $p(t) = O(A)$. Hence, there will be a time window $\mathcal{\tau}$ such that $\sigma < p(t) < A$ for $t \in \mathcal{\tau}$ in which the noise accumulates prior to the appearance of a large deviation.
The system begins to follow Eqs. \eqref{x}-\eqref{xp} at $t=t_0$ and after $t-t_0>T$ the momenta behave as
\begin{equation}\begin{split}
p(t)=p(t_0)\exp\left [-\int_{t_0}^{t}dsF'(x(s),s)\right ]\simeq p(t_0)e^{-\lambda_s (t-t_0)},
\label{pt}
\end{split}\end{equation}
where $\lambda_s < 0$ is the Lyapunov exponent of the stable periodic orbit defined as $\lambda_s=\frac{1}{T}\int_{t}^{t+T}dz F(x_s(z),z)$. Therefore, during a ``warning time'' $\tau_W \sim -\frac{1}{\lambda_S}\ln\left[\frac{A}{p(t_0)}\right]$, Eq. (\ref{pt}) describes the noise accumulation before $x(t)$ exhibits a transition.  Clearly, because 
the warning time is inversely proportional to $\lambda_s < 0$, in less stable orbits we can determine a jump precursor earlier.

\begin{figure}
	\centering
	\includegraphics[width=0.49 \textwidth]{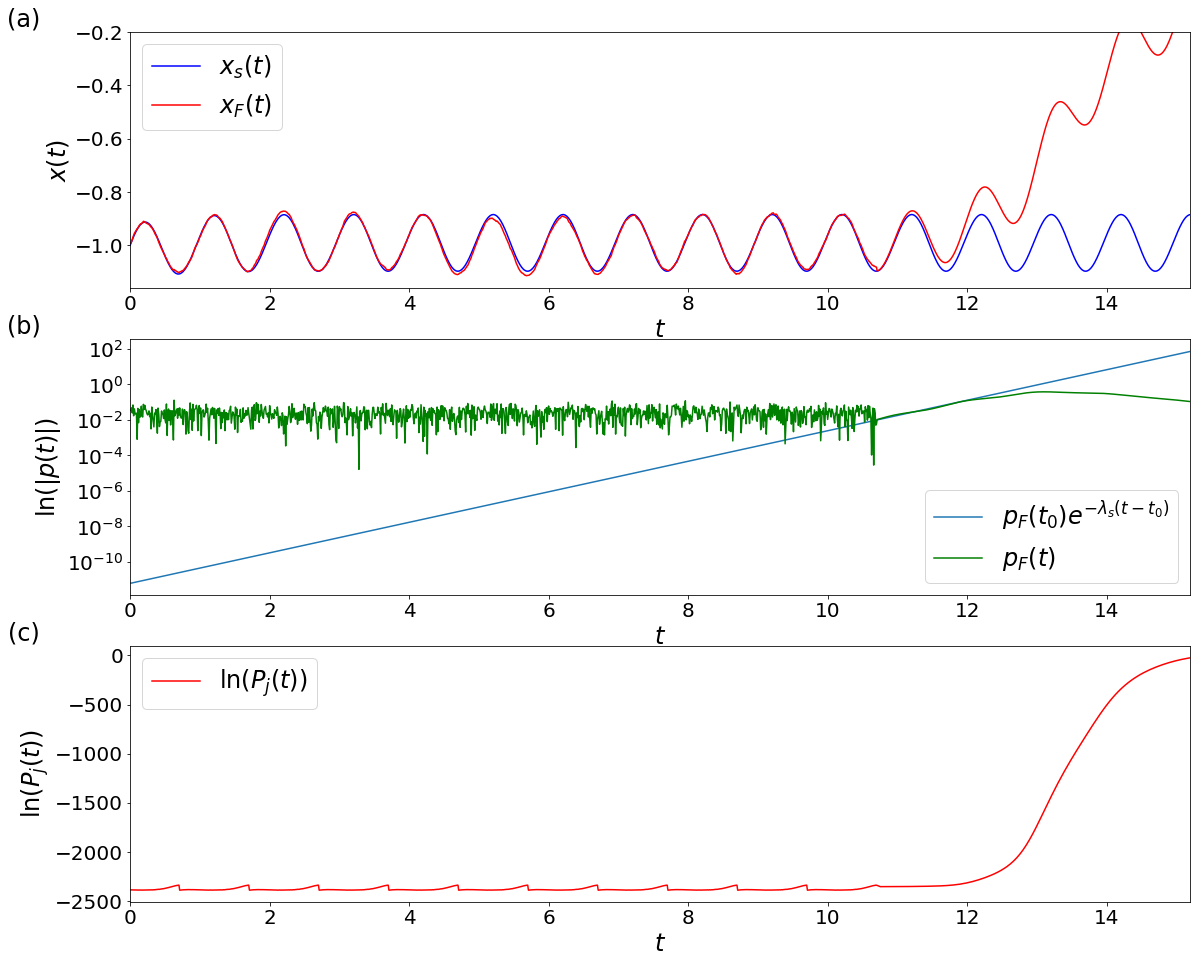}
	\caption{(a) Time evolution of the position $ x_F(t) $, compared to the stable periodic orbit $x_S(t)$. (b) A semi-log plot of $ |p(t)| $. (c) Time behavior of the jump probability computed from Eq. (\ref{probbb}).}
	\label{F3}
\end{figure}

The rare event momentum precursor is demonstrated numerically in Fig. \ref{F3}, which shows the transition from one potential well to the other.  We used the instanton dynamics described in \S 4 to force the appropriate accumulation of noise in the case where $F(x,t)= x-x^3+0.7\cos(2\pi t)$ and $\sigma=0.01$ in Eq. (\ref{lang}), which we evolve for ten periods  with initial condition $x(0)=-1$.
We then use Eqs. \eqref{x}-\eqref{boundary} to simulate the jump. Comparing Figs. \ref{F3}(a) and (b) one observes the ``momentum anticipation'' of the deviation of the trajectory $x_F(t)$ from the stable periodic orbit $x_s(t).$

We can obtain an accurate estimate of the time interval in which the jump will occur from Eqs. \eqref{x}, \eqref{xp},\eqref{probbb} and \eqref{S}.  Within each period there are only a few highly probable paths and thus upon observation of optimal noise accumulation, we can determine which path the system is following. Hence, we can estimate the corresponding jump time, $ t_j $, when the system shifts to the other stable basin (Fig. \ref{Sp0} (b)).  Therefore, by studying the fluctuations around the stable periodic orbit to determine when they begin to behave as $p(t)$, we can predict if the system is approaching a jump by computing the jump probability and time. Next we provide a systematic outline of our prediction strategy. 

\section{4. Prediction Scheme}
\label{sect_sum}

\begin{figure}
	\centering
	\includegraphics[width=0.48\textwidth]{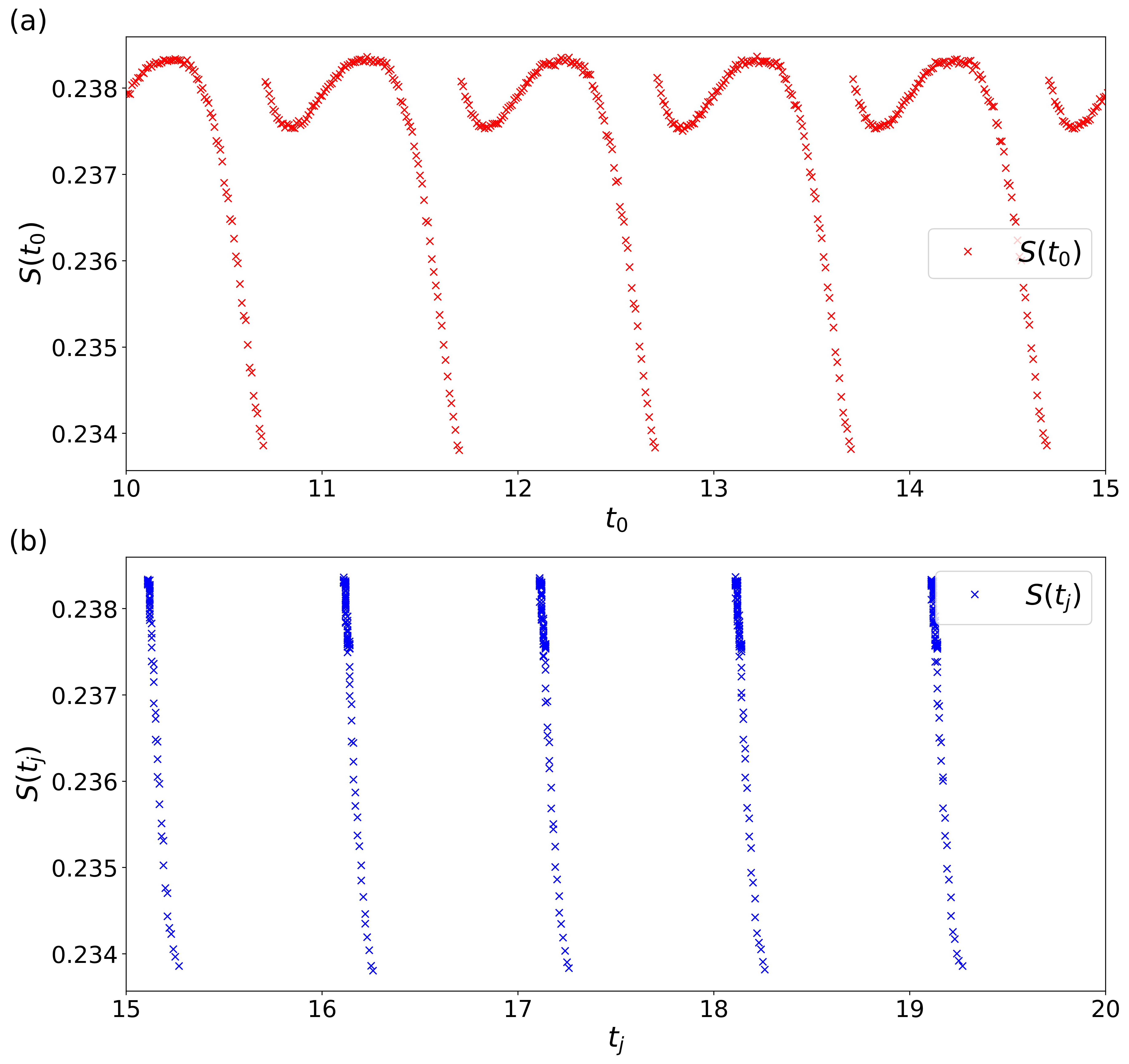}		
\caption{Values of the action corresponding to different values of (a) $ t_0 $ and (b) $ t_j $. The system has been evolved for 10 periods setting $ p(t)=0 $ in order to be sure that it follows the stable periodic orbit, after which Eq. (\ref{xp}) was used for $p(t)$.}
	\label{Sp0}
\end{figure}

Our program for the prediction and study of rare events in stochastic resonance consists of the following five main steps. 
\begin{itemize}
	\item[(1)] We start with the nonautonomous Langevin equation Eq. (\ref{lang}) describing the time evolution of the system,
	\begin{equation}
	\dot{x}(t)=F(x(t),t)+\sqrt{2}\sigma \xi(t), \nonumber
	\end{equation}
	assuming we know $ F(x(t),t)$ and $ \sigma $. See \cite{moon2017unified} and \cite{hasselmann1976stochastic} regarding the construction of these expressions from the data (the latter paper treating the autonomous case). 
	\item[(2)] We determine the instantons as follows. First, we evolve 
\begin{equation}
\dot{x}_s(t) =F(x_s(t),t), \nonumber
\label{start}
\end{equation}
	with $ x_s(t=0) $ chosen inside one potential well. After an initial transient, the system state evolves following the stable periodic orbit. When this condition is satisfied at $ t=t_0 $, we modify the earlier equation to
	\begin{align} 
\dot{x}(t) &=2p(t)+F(x(t),t),  \nonumber \\
\dot{p}(t) &=-p(t)F'(x(t),t), \nonumber
\end{align}
with  
\begin{equation}
x(t_0)=x_s(t_0) \qquad \text{and} \qquad p(t_0)=p_0 > \sigma.
\label{boundary0}
\end{equation}
We evolve the system many times until we observe a shift to another stable basin at $ t=t_j $, using the same value of $ p_0 $ but with different values of $ t_0 $ chosen inside one period. For each of these paths we compute the relative action 
\begin{equation}
S[x_k(t)]=\int_{t_0}^{t_j}p_k^2(t) dt. \nonumber \\
\end{equation} 
Because $S(t_0)$ is periodic, there will be only one instanton in every period.
We find the instanton as the path that minimizes the action in every period (See Fig. \ref{Sp0}).
	\item[(3)] We isolate the noise from the data using 
	the Langevin equation as in Eq. (\ref{noise_acc}); 
	\begin{equation}
	\xi(t)= \frac{1}{\sqrt{2} \sigma}[\dot{x}(t)-F(x(t),t)], \nonumber
	\end{equation}
	which is related to the conjugate momentum as $p(t)=\sigma \xi(t)/\sqrt{2} $. 
	\item[(4)] Prior to the jump the conjugate momentum is expected to increase exponentially as $p(t)=p_0e^{-\lambda_s (t-t_0)}$.  Thus, we scrutinize the behavior of $ p(t) $ obtained from the data of $x(t)$ through Eq. (\ref{noise_acc}) until it ceases to exhibit fluctuations of order $ \sigma $ near zero and begins to grow.  We compare its behavior to the conjugate momentum of the instanton and if the noise structure differs from optimality we are unable to make predictions; because the noise structure is not optimal, the jump is more rare.  However, when the noise structure is optimal, we can accurately estimate the jump probability and time.
\end{itemize}

Next we demonstrate this scheme in a numerical example.

\section{5. Numerical Demonstration}
\label{sect_res}

In order to demonstrate this strategy, we evolve Eq. (\ref{lang}) numerically for a very long time, until the system jumps from one potential well to the other. We use a quartic potential $ U(x)=-x^2/2+x^4/4 $ with $ A=0.7 $, $ \omega=2\pi $ and $ \sigma=0.0727 $.
These parameters are chosen to maximize the difference between $ A $ and $ \sigma $ and yet still yield a jump in a tractable	simulation time.

In the absence of noise and periodic forcing, the resulting Langevin equation has two stable periodic solutions separated by an unstable one. We apply our prediction scheme to study the transition between the two stable periodic solutions in the regime $\sigma < A$. The results are discussed in the following.
 
\begin{figure}
	\centering
	\includegraphics[width=0.48\textwidth]{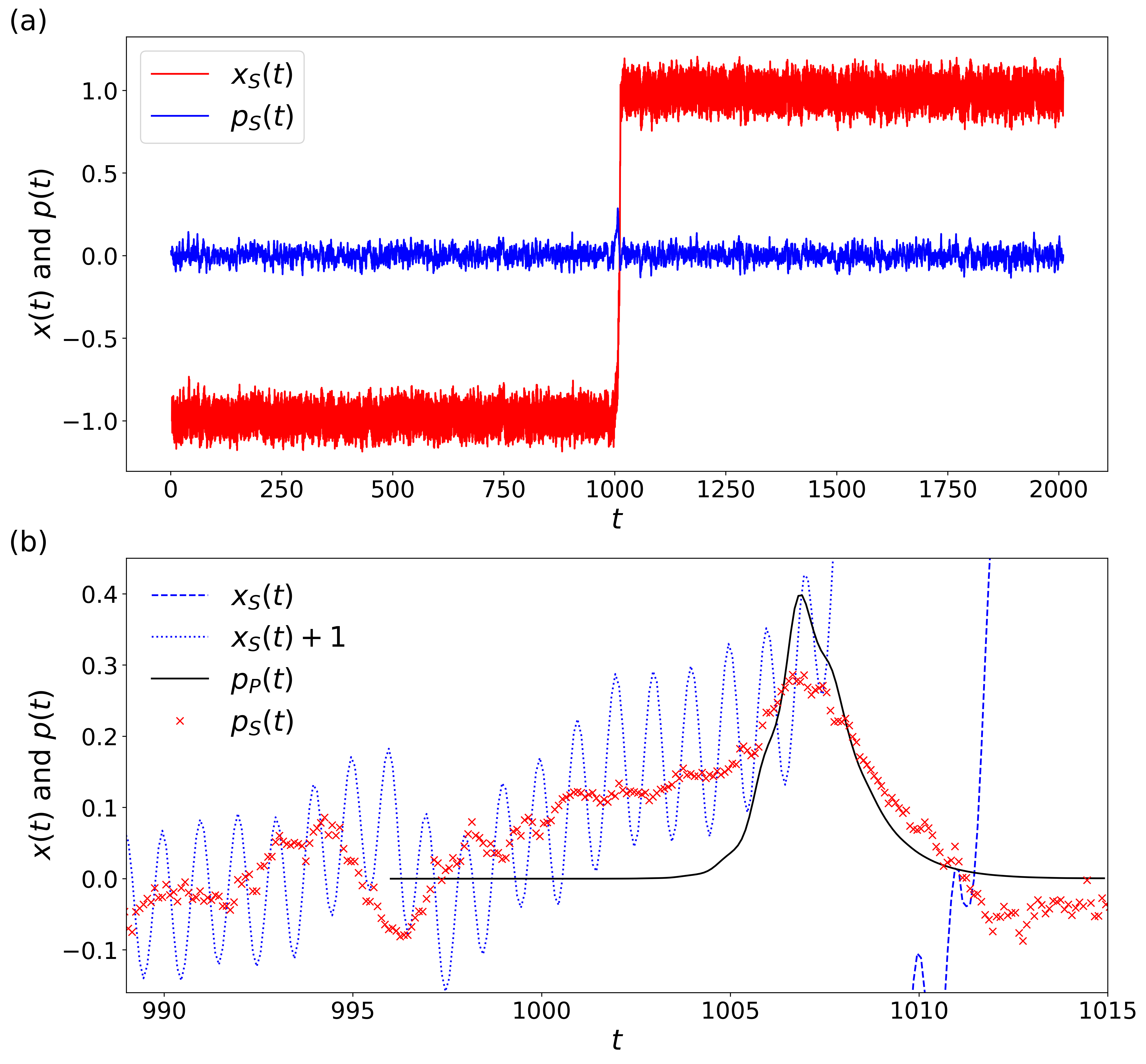}
	\caption{(a) Plot of the position of the system $x(t)$  and of $ p(t)=\frac{1}{2}[\dot{x}(t)-F(x(t),t)]  $ computed numerically (using a leap frog algorithm) from Eq. (\ref{lang}), and denoted with the subscript $ S $.  (b) Expansion of the previous plot and comparison with $ p(t) $ computed from  Eqs. (\ref{x})-(\ref{xp}), denoted with the subscript $ P $. Note that  $x(t)$ first hits the origin at approximately  $t=1011.5$, which is preceded with a peak in $p(t)$ by a time of 4 periods. The data are smoothed using a moving average.} 
	\label{F1}
\end{figure}

\begin{figure}
	\centering
	\includegraphics[width=0.48\textwidth]{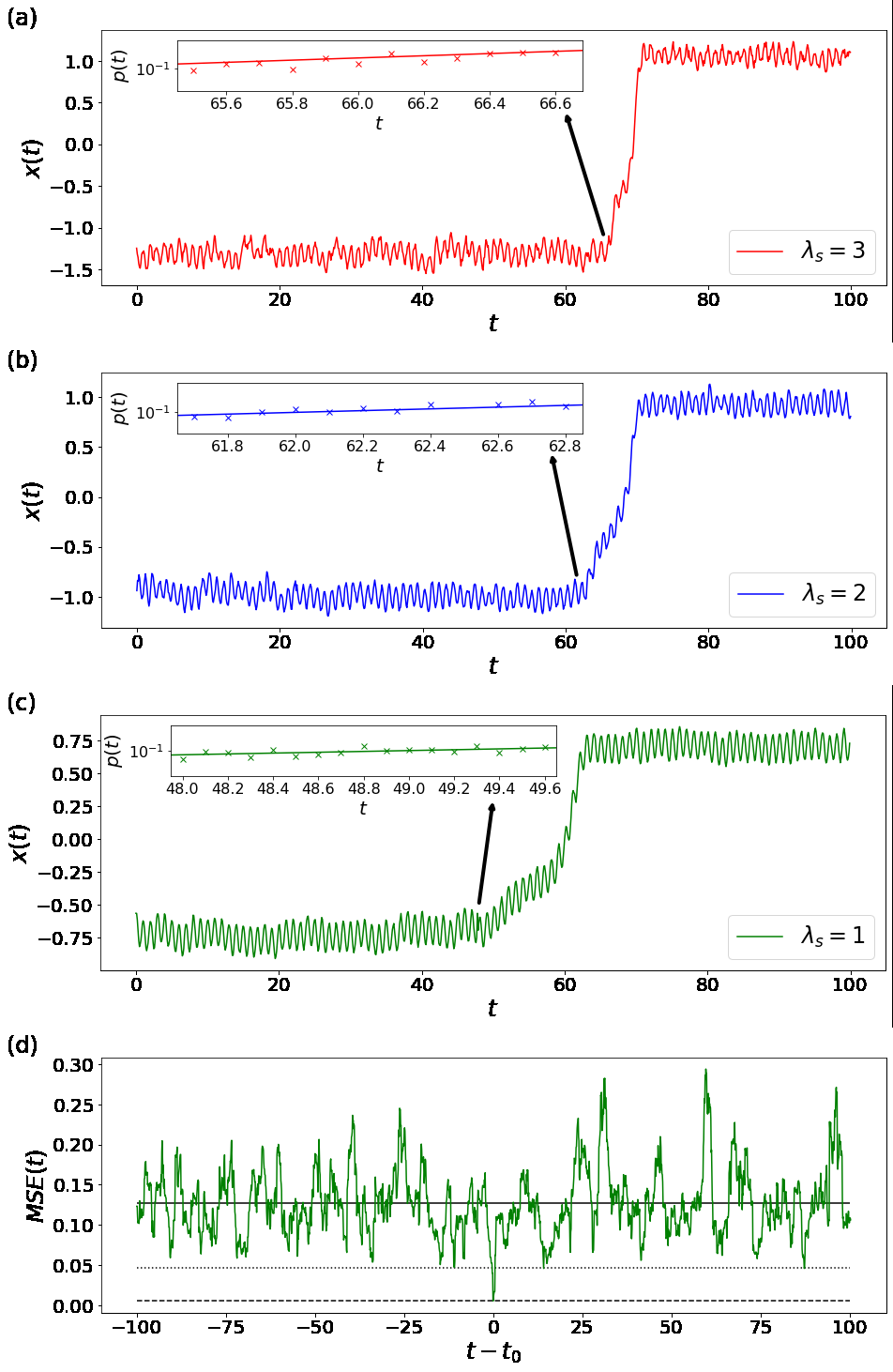}
	\caption{The trajectories of $x(t)$ for 100 periods and $p(t)$ immediately prior to the jump (along with the analytical prediction)  for three different simulations associated with three different Lyapunov exponents are shown in (a)--(c).  
	(d) The MSE of the noise realizations with respect to the optimal behavior predicted in Eq. (\ref{pt}), using the data from the simulation shown in (c) displayed over a larger time window. The solid line is the average MSE, the dashed line is the MSE corresponding to the values of $ p(t) $ in the time window highlighted in the top left of (c) and the dotted line is the minimum value reached by the MSE if these values of $ p(t) $ are removed.} 
	\label{F2}
\end{figure}

Fig. \ref{F1}(a) shows the time evolution of the system state $x(t)$ and of the deviation $ p(t) $ from deterministic flow (the noise) over a time of 2000 periods. We observe that $ p(t) $ exhibits small oscillations around zero, showing a peak near the jump. Since the system is driven by white noise, $ p(t) $ has no temporal structure in the time frame considered, save for the increment very near the jump. Indeed, as described above, this jump can only be observed if the noise accumulates in an optimal way.
 Namely, in this region the noise {\em does not behave randomly}.  Rather, in order to drive the system to a jump, the noise should form a specific structure  that depends on the shape of the potential. 
 
Fig. \ref{F1}(b) shows in detail the behavior of $ p(t) $ close to the jump. When the value of $ p(t) $ is near zero, we expect that the noise, and thus $ p(t) $, will accumulate randomly. Importantly, in that region, because there are many ways for the noise to accumulate with equal probability, finding an optimal path is meaningless. However, when $ p(t) $ deviates from zero we find a substantially different cumulative influence of the noise. Namely, because the probability differences between paths increase exponentially, only a single path becomes relevant. This path is very near the optimal path satisfying Eqs. (\ref{x})-(\ref{xp}). Indeed, the solution of Eq. (\ref{xp}) and numerical shapes of $ p(t) $ agree well near the transition; increasing exponentially rather far from zero as $ e^{-\lambda_s t} $ and then, when approaching the unstable periodic orbit around the maximum of the potential, it begins to decrease as $ e^{-\lambda_u t} $, with $ \lambda_u $ the Lyapunov exponent of the unstable periodic orbit defined as $\lambda_u=\frac{1}{T}\int_{t}^{t+T}dz F(x_u(z),z)$. Clearly the unstable periodic orbit will not persist and, after reaching the maximum of the potential, the system will immediately fall into the stable periodic orbit around the minimum of the other well. However, the asymmetric influence of the noise as the system transitions is responsible for the deviation of the numerical and the analytical prediction near the peak.

Figs. \ref{F2}(a)-(c) show $x(t)$ in three different simulations and $p(t)$ during the interval in which noise is accumulating shortly before the transition.  In each simulation we have modified the shape of the potential in order to change the value of the Lyapunov exponent of the stable orbit and the noise amplitude to make the waiting time for the jump comparable in each simulation. This is accomplished by varying the value of $a$ in ${-U^\prime(x)}=a\,x-x^3$, with $a$ = 1.5, 1 and 0.5 in Figs. \ref{F2}(a)-(c) respectively, as well as the values of $\sigma$, with $ \sigma$=0.1233, 0.0727 and 0.0632 in Figs. \ref{F2}(a)-(c) respectively. In all cases, $A = 0.7$ and $\omega = 2\pi$. These figures demonstrate the optimal nature of the noise accumulation described by the instanton Eqs. (\ref{x}), (\ref{xp}).
In Fig. \ref{F2} (d) we show the mean standard error (MSE) between the realizations of $ p(t) $ constructed from the realizations of the noise and its optimal behavior described in Eq. (\ref{pt}).  This is defined as 
\begin{equation}
MSE(t)=
\dfrac{1}{n}\min_{p_0}\left\{\sum_{i=m}^{m+n}\left(p_i-p_0e^{\lambda_s t_i}\right)\right\}.
\end{equation}
We computed the MSE using the noise from Fig. \ref{F2} (c) taken over a 200 period time window spanning an interval exhibiting linear behavior on a logarithmic scale.  We used  a moving time window of size $n=22 =2.2\, dt (\sim\tau_W) $, in each of which we chose the value of $ p_0 $ that minimizes the MSE between the values of $p(t)$ constructed from the realizations of the noise and the exponential slope given in Eq. (\ref{pt}). We find that the MSE assumes larger values--ten times larger at least--for for the time interval in which the noise deviates from $ p_0e^{-\lambda_s t}$.  Thus, these simulations show that the time window during which the noise acquires a specific structure increases thereby decreasing the value of the Lyapunov exponent, consistent with the discussion in \S 2.  Namely, a 
smaller Lyapunov exponent implies a slower optimal accumulation of the noise. Therefore, the system will take more time to shift to another stable basin so that the warning time of a rare event increases for less stable potentials.

When $ \sigma \ll A $ and $ \sigma \ll \Delta U$ the jump probability is extremely small and thus poses a substantial numerical challenge.  Having shown that our strategy works well under less extreme cases we have thereby quantified how the conjugate momentum of the instanton organizes the noise prior to the jump.  Therefore, we expect that the same noise organization process will be operative in the case where $ \sigma \ll A $ and $ \sigma \ll \Delta U $ and the system behavior shown in Fig. \ref{F3} will be recovered.

\section{6. Conclusion}
\label{sect_concl}

We have developed a theory to study and find precursors to noise-induced rare events within the general framework of stochastic resonance.  
In stochastic resonance, a periodically and noise forced system in a double-well potential jumps between minima, but the time-scale separation of these forcings insures that the system oscillates for a long time about one of the local minima of the potential and only very rarely jumps to the other minima.  The ubiquity of such transitions underlies the importance of trying to predict when they will occur. 

We have used a path-integral method to determine the particular manner in which the fluctuations around the unperturbed deterministic flow must organize prior to the system jump. We have showed how to predict the time within a period when the system will transit to another minima, and have harnessed the signature of this fluctuation behavior as an advanced indicator of a potential jump, as well as computing the probability of such rare events.  The method provides a framework to examine data in a manner that facilitates predictions across a broad spectrum of stochastic systems. 
Finally, the approach identifies a short well defined structure immediately prior to the rare event.  The detection of such structures in a prediction setting is a central aspect of many machine learning approaches to rare event predictions \cite[e.g.][]{Lim2019}, which provide a test bed for a wide range of approaches.

\acknowledgments
The authors acknowledge the support of Swedish Research Council grant no. 638-2013-9243.  

\appendix

\section{Derivation of the Most Probable Path}

%\section*{Derivation of the Most Probable Path}

%This appendix is devoted to briefly illustrating 
Here we describe the steps in the derivation of the results of Section \S \ref{sect_outline}. This material may be unnecessary for readers who are familiar with this general path integral formulation.  Equation numbers not adorned with the capital letter \textcolor{red}{A} refer to those in the main text. 

The starting point is the conditional probability density, $ p(x_f,t_f|x_i,t_i) $, that the process $x(t)$ reaches the point $x_f$ at time $t_f$, given that it started at the point $x_i$ at time $t_i$. The evolution of this probability density is governed by the Fokker-Planck equation 
\begin{equation}
\frac{\partial}{\partial t} \,p(x,t)=\frac{\partial}{\partial x}\, \left[ \left(-F(x,t)+\sigma^2\frac{\partial}{\partial x} \right) p(x,t) \right], 
\tag{A.1}
\label{eq:FPE}
\end{equation}
with the boundary conditions
\begin{equation}
x(t_i)=x_i, \qquad \text{and} \qquad x(t_f)=x_f.
\tag{A.2}
\label{eq:FPEb}
\end{equation}
In order to solve the Fokker-Plank equation, we rewrite Eq. (1) as the following system of two It\^o stochastic differential equations; 
\begin{align}
&dy=\sqrt{2}\sigma dW \tag{A.3}\label{sistem1}\qquad \text{and}\\
&dx=F(x,t)dt+\sqrt{2}\sigma dW,
\tag{A.4}
\label{sistem2}
\end{align}	
where $W$ is a Wiener process and thus  $\xi(t) =\frac{dW}{dt}$, a formal time-derivative of $W$, which is 
Gaussian white noise with zero mean and correlation function given by Eq. (3).

We consider now a uniform discretization of time in the interval $ [t_{i},t_{f}] $ with increment $ \Delta t=(t_{f}-t_{i})/N $ and we define the random vectors $ x=(x_0,x_1,...,x_N) $ and $ y=(y_0,y_1,...,y_N) $, where $x_k = x(t_{i}+k\Delta t)$, $y_k = y(t_{i} + k \Delta t)$, for $k=0,1,\dots,N$. The probability density, $ \rho_y(y)$, of $ y $ can be written as
\begin{equation}
\rho_y(y)=\left(\frac{1}{4\pi\sigma^2\Delta t}\right)^{N/2}\exp\left[\sum_{n=0}^{N-1}\dfrac{(y_{n+1}-y_n)^2}{4 \sigma^2\Delta t}\right].
\tag{A.5}
\end{equation}
We integrate Eqs. (\ref{sistem1}, \ref{sistem2}) to find
\begin{align}
&y_n=\sqrt{2}\sigma \int_{t_{i}}^{t_n}dW \qquad \text{and} \qquad \tag{A.6}\\
&x_n=\int_{t_{i}}^{t_n}F(x,t)dt+\sqrt{2}\sigma \int_{t_{i}}^{t_n} dW,
\tag{A.7}
\label{sistemI}
\end{align}	
for $n=1,2,\dots,N$. 
Using the composite trapezoidal rule, it follows that
\begin{equation}
y_n=x_n-\dfrac{\Delta t}{2}\sum_{m=0}^{n-1}(F(x_m,t_m)+F(x_{m+1},t_{m+1}))+O[(\Delta t)^2].
\tag{A.8}
\end{equation}
Differentiating with respect to $ x_n $ we find 
\begin{equation}\begin{split}
\dfrac{dy_n}{dx_n}&=1-\dfrac{\Delta t}{2}F'(x_n,t_n)+O[(\Delta t)^2]\\
&= \exp\left[-\dfrac{\Delta t}{2}F'(x_n,t_n)\right]+O[(\Delta t)^2],
\end{split}
\tag{A.9}
\end{equation}
which is the Jacobian of the transformation allowing us to construct the probability density $ \rho_x(x) $ starting from $ \rho_y(y) $. It follows that $ \rho_x(x) $, neglecting the $ O[(\Delta t)^2]$ terms, is
\begin{equation}\begin{split}
&\rho_x(x)=\left(\frac{1}{4\pi\sigma^2\Delta t}\right)^{N/2}\exp\left[-\dfrac{\Delta t}{2}\sum_{n=1}^{N}F'(x_n,t_n)\right] \\
&\ \  \exp\bigg[ -\sum_{n=0}^{N-1} \\
&\ \ \dfrac{\left(x_{n+1}-x_n-\dfrac{\Delta t}{2}(F(x_n,t_n)+F(x_{n+1},t_{n+1}))\right)^2}{4\sigma^2\Delta t}\bigg].
\end{split}
\tag{A.10}
\end{equation}
Taking the limit $ \Delta t \to 0 $ we obtain the following formal expression for a measure in the $ x $ variable (neglecting the subleading term of order $ \sigma^2 $ in the exponential)
\begin{equation}\begin{split}
d\mu_x&=\rho_x(x)\prod_{n=1}^{\infty}dx_n\\
&=\dfrac{1}{Z}\exp\left[-\frac{1}{4\sigma^2}\int_{t_i}^{t_f}(\dot{x}(t)-F(x(t),t))^2 dt \right]\prod_{n=1}^{\infty}dx_n,
\label{W} 
\end{split}
\tag{A.11}
\end{equation}
where $ Z $ is the appropriate normalization factor. We emphasize that the above representation of $d\mu_x$ as a Lebesgue measure is a heuristic definition and is commonly used in the physics literature, despite the fact that a Lebesgue measure cannot be defined on the (infinite-dimensional) space of continuous functions. However, when performing formal derivations it is convenient to treat the measure as if it has a density with respect to a Lebesgue measure.

In view of this, we denote $ \prod_{n=1}^{\infty}dx_n/Z=\arrowvert dx(t) \arrowvert $ and using Eq. (\ref{W}) we can write the solution of Eq. (\ref{eq:FPE}) with the boundary conditions (\ref{eq:FPEb}) as
\begin{align}
&P(x_f,t_f|x_i,t_i)=\int_{x(t_i)\equiv x_i}^{x(t_f) \equiv x_f}\arrowvert dx(t)\arrowvert \exp\left \{-\dfrac{1}{\sigma^2}S[x(t)]\right \} \nonumber \\
&\hspace{3cm} +O(\sigma^2),
\tag{A.12}
\label{prob2}
\end{align}	
where $S[x(t)]$ is the action functional given by
\begin{equation}
S[x(t)]=\int_{t_i}^{t_f}L[x(t),\dot{x}(t),t]dt,
\tag{A.13}
\label{s}
\end{equation}
with the Lagrangian
\begin{equation}
L[x(t),\dot{x}(t),t]=\dfrac{1}{4}[\dot{x}(t)-F(x(t),t)]^2. 
\tag{A.14}
\label{l}
\end{equation}	
\begin{comment}
This result can be rigorously justified using the sample path large deviation theory of Freidlin-Wentzell \cite{freidlin2012random}. 
\end{comment}

We use the results \eqref{prob2}-\eqref{l} to derive the optimal path, i.e., the most likely path through which the system state passes between two given points.  In order to provide the overall structure our derivation is heuristic, but the reader is referred to [22] for a rigorous approach and discussion.   Because $ \sigma \ll 1 $, the path integral in Eq. (\ref{prob2}) is dominated by the local minima of the action, and thus we integrate only over those paths that satisfy
\begin{equation}
\frac{\delta S[x_k(t)]}{\delta x_k(t)} \bigg|_{x_{k}(t)\equiv x_k^*(t)}=0,
\tag{A.15}
\label{szero}
\end{equation}
where $\frac{\delta S}{\delta x_k(t)}$ denotes the functional derivative of $S$ with respect to $x_k$ at time $t$, here evaluated at
\begin{equation}
x^*_k(t_i)=x_i \qquad \text{and} \qquad x^*_k(t_f)=x_f.
\tag{A.16}
\label{boundary}
\end{equation}
The $x_k^*(t)$ are  the local minimizers of the action, defining the {\it most probable} or {\it optimal paths} connecting the points $(x_i, t_i)$ and $(x_f, t_f)$, also known as {\it instantons}.
Using Eqs. (\ref{s})-(\ref{boundary}), we find that $ x^*_k(t) $ satisfies the following second order differential equation:
\begin{equation}
\ddot{x}^*_k(t)=\dot{F}(x^*_k(t),t)+F(x^*_k(t),t)F'(x^*_k(t),t),
\tag{A.17}
\end{equation} 
where the dot (prime) denotes differentiation with respect to time (position).

It is convenient to represent this boundary value problem using a Hamiltonian description, with the Hamiltonian:
\begin{equation}
H(x_k,p_k,t)=p_k \dot{x}_k -L=p_k^2+p_k F(x_k,t),
\tag{A.18}
\end{equation}  
where we have dropped the superscript $*$, and the ``momentum'' variable $p_k(t) = \frac{1}{2}[\dot{x}_k(t)-F(x_k(t),t)]$ measures  deviation from the deterministic unperturbed flow. The Hamiltonian variables satisfy the system of first order differential equations Eqs. (4, 5).
Taking into account only these optimal paths and neglecting the $O(\sigma^2)$ term, the expressions given in Eqs. (8, 9) follows.

\end{document}